\input harvmac.tex

\input epsf.tex

\def\fig#1#2#3{
\par\begingroup\parindent=0pt\leftskip=1cm\rightskip=1cm\parindent=0pt
\baselineskip=11pt
\global\advance\figno by 1
\midinsert
\epsfxsize=#3
\centerline{\epsfbox{#2}}
\vskip 12pt
{\bf Fig.\ \the\figno: } #1\par
\endinsert\endgroup\par
}

\def\figin{\epsfcheck\figin}\def\figins{\epsfcheck\figins}
\def\epsfcheck{\ifx\epsfbox\UnDeFiNeD
\message{(NO epsf.tex, FIGURES WILL BE IGNORED)}
\gdef\figin##1{\vskip2in}\gdef\figins##1{\hskip.5in}
\else\message{(FIGURES WILL BE INCLUDED)}%
\gdef\figin##1{##1}\gdef\figins##1{##1}\fi}
\def\DefWarn#1{}
\def\figinsert{\goodbreak\midinsert}
\def\ifig#1#2#3{\DefWarn#1\xdef#1{fig.~\the\figno}
\writedef{#1\leftbracket fig.\noexpand~\the\figno}%
\figinsert\figin{\centerline{#3}}\medskip\centerline{\vbox{\baselineskip12pt
\advance\hsize by -1truein\noindent\footnotefont{\bf Fig.~\the\figno:} #2}}
\bigskip\endinsert\global\advance\figno by1}




{ \Title
{\vbox{\baselineskip12pt \hbox{}}}
{\vbox{
{\centerline {A silence black hole: Hawking radiation}
{\centerline {at the Hagedorn temperature}
}}}}

\bigskip
\centerline{O. Lorente-Espin and P. Talavera }
\bigskip~
 
\centerline{ Department de F{\'\i}sica i
Enginyeria Nuclear,}
\centerline{ Universitat Polit\`ecnica de Catalunya,
Comte Urgell, 187, E-08036 Barcelona, Spain}

\medskip



\vskip .3in

\baselineskip12pt
We compute semi-classically the Hawking emission for different types of black hole in 
type II string theory.
In particular we analyze the thermal transition between NS5 branes and Little String Theory, finding  compelling evidence for information recovering.
We find that once the near horizon limit is taken the emission of a full family of models is exactly
{\sl thermal} even if back-reaction is taken into account. 
Consequently these theories are non-unitary and can not convey any information about the black hole internal states. It is argue that this behaviour matches the string theory expectations. 

\vfill

\noindent
\Date{March 2008}


\eject

\lref\AharonyTT{
  O.~Aharony and T.~Banks,
  ``Note on the quantum mechanics of M theory,''
  JHEP {\bf 9903}, 016 (1999)
  [arXiv:hep-th/9812237].
}

\lref\KutasovUA{
  D.~Kutasov and N.~Seiberg,
  ``Noncritical superstrings,''
  Phys.\ Lett.\  B {\bf 251}, 67 (1990).
}

\lref\HawkingFD{
  S.~W.~Hawking and G.~T.~Horowitz,
  ``The Gravitational Hamiltonian, action, entropy and surface terms,''
  Class.\ Quant.\ Grav.\  {\bf 13}, 1487 (1996)
  [arXiv:gr-qc/9501014].
}

\lref\MinwallaXI{
  S.~Minwalla and N.~Seiberg,
  ``Comments on the IIA NS5-brane,''
  JHEP {\bf 9906}, 007 (1999)
  [arXiv:hep-th/9904142].
}

\lref\NarayanDR{
  K.~Narayan and M.~Rangamani,
  ``Hot little string correlators: A view from supergravity,''
  JHEP {\bf 0108}, 054 (2001)
  [arXiv:hep-th/0107111].
}

\lref\BerkoozMZ{
  M.~Berkooz and M.~Rozali,
  ``Near Hagedorn dynamics of NS fivebranes, or a new universality class  of
  coiled strings,''
  JHEP {\bf 0005}, 040 (2000)
  [arXiv:hep-th/0005047].
}

\lref\KapustinCI{
  A.~Kapustin,
  ``On the universality class of little string theories,''
  Phys.\ Rev.\  D {\bf 63}, 086005 (2001)
  [arXiv:hep-th/9912044].
}

\lref\StromingerSH{
  A.~Strominger and C.~Vafa,
  ``Microscopic Origin of the Bekenstein-Hawking Entropy,''
  Phys.\ Lett.\  B {\bf 379}, 99 (1996)
  [arXiv:hep-th/9601029].
}

\lref\CallanAT{
  C.~G.~Callan, J.~A.~Harvey and A.~Strominger,
  ``Supersymmetric string solitons,''
  arXiv:hep-th/9112030.
}

\lref\AharonyUB{
 O.~Aharony, M.~Berkooz, D.~Kutasov and N.~Seiberg,
 ``Linear dilatons, NS5-branes and holography,''
 JHEP {\bf 9810}, 004 (1998)
 [arXiv:hep-th/9808149].
}

\lref\hoo{
 K.~Hori, H.~Ooguri and Y.~Oz,
   ``Strong coupling dynamics of four-dimensional N = 1 gauge theories from  M
  theory fivebrane,''
  Adv.\ Theor.\ Math.\ Phys.\  {\bf 1}, 1 (1998)
  [arXiv:hep-th/9706082].
}

\lref\BuchelDG{
  A.~Buchel,
  ``On the thermodynamic instability of LST,''
  arXiv:hep-th/0107102.
}

\lref\KutasovJP{
  D.~Kutasov and D.~A.~Sahakyan,
  ``Comments on the thermodynamics of little string theory,''
  JHEP {\bf 0102}, 021 (2001)
  [arXiv:hep-th/0012258].
}

\lref\MaldacenaYA{
 J.~M.~Maldacena,
 ``Statistical Entropy of Near Extremal Five-branes,''
 Nucl.\ Phys.\  B {\bf 477}, 168 (1996)
 [arXiv:hep-th/9605016].
}

\lref\HarmarkHW{
 T.~Harmark and N.~A.~Obers,
 ``Hagedorn behaviour of little string theory from string corrections to
 NS5-branes,''
 Phys.\ Lett.\  B {\bf 485}, 285 (2000)
 [arXiv:hep-th/0005021].
}

\lref\ParikhMF{
  M.~K.~Parikh and F.~Wilczek,
  ``Hawking radiation as tunneling,''
  Phys.\ Rev.\ Lett.\  {\bf 85}, 5042 (2000)
  [arXiv:hep-th/9907001].
}

\lref\Birkhoff{
  G.~D.~Birkhoff,
  ``Relativity and Modern Physics,''
  Harvard University Press, Cambridge (1927).
}

\lref\ParikhIH{
  M.~K.~Parikh,
  ``A secret tunnel through the horizon,''
  Int.\ J.\ Mod.\ Phys.\  D {\bf 13}, 2351 (2004)
  [Gen.\ Rel.\ Grav.\  {\bf 36}, 2419 (2004)]
  [arXiv:hep-th/0405160].
}

\lref\HawkingSW{
  S.~W.~Hawking,
  ``Particle Creation By Black Holes,''
  Commun.\ Math.\ Phys.\  {\bf 43}, 199 (1975)
  [Erratum-ibid.\  {\bf 46}, 206 (1976)].
}

\lref\HawkingRA{
  S.~W.~Hawking,
  ``Breakdown Of Predictability In Gravitational Collapse,''
  Phys.\ Rev.\  D {\bf 14}, 2460 (1976).
}

\lref\PreskillTB{
  J.~Preskill, P.~Schwarz, A.~D.~Shapere, S.~Trivedi and F.~Wilczek,
  ``Limitations on the statistical description of black holes,''
  Mod.\ Phys.\ Lett.\  A {\bf 6}, 2353 (1991).
}

\lref\HawkingTU{
  S.~W.~Hawking,
  ``Gravitational radiation from colliding black holes,''
  Phys.\ Rev.\ Lett.\  {\bf 26}, 1344 (1971).
}

\lref\CallanHS{
  C.~G.~Callan, R.~C.~Myers and M.~J.~Perry,
  ``Black Holes in String Theory,''
  Nucl.\ Phys.\  B {\bf 311}, 673 (1989).
}

\lref\MitraQA{
  P.~Mitra,
  ``Hawking temperature from tunnelling formalism,''
  Phys.\ Lett.\  B {\bf 648}, 240 (2007)
  [arXiv:hep-th/0611265].
}

\lref\ArzanoRS{
  M.~Arzano, A.~J.~M.~Medved and E.~C.~Vagenas,
  ``Hawking radiation as tunneling through the quantum horizon,''
  JHEP {\bf 0509}, 037 (2005)
  [arXiv:hep-th/0505266].
}

\lref\GubserEG{
  S.~S.~Gubser, A.~A.~Tseytlin and M.~S.~Volkov,
  ``Non-Abelian 4-d black holes, wrapped 5-branes, and their dual descriptions,''
  JHEP {\bf 0109}, 017 (2001)
  [arXiv:hep-th/0108205].
}

\lref\KutasovJP{
  D.~Kutasov and D.~A.~Sahakyan,
  ``Comments on the thermodynamics of little string theory,''
  JHEP {\bf 0102}, 021 (2001)
  [arXiv:hep-th/0012258].
}

\lref\KernerVU{
  R.~Kerner and R.~B.~Mann,
  ``Tunnelling, temperature and Taub-NUT black holes,''
  Phys.\ Rev.\  D {\bf 73}, 104010 (2006)
  [arXiv:gr-qc/0603019].
}

\lref\UnruhDB{
  W.~G.~Unruh,
  ``Notes on black hole evaporation,''
  Phys.\ Rev.\  D {\bf 14}, 870 (1976).
}

\lref\CaseroPT{
  R.~Casero, C.~Nunez and A.~Paredes,
  ``Towards the string dual of N = 1 SQCD-like theories,''
  Phys.\ Rev.\  D {\bf 73}, 086005 (2006)
  [arXiv:hep-th/0602027].
}

\lref\ParikhRH{
  M.~K.~Parikh,
  ``Energy conservation and Hawking radiation,''
  arXiv:hep-th/0402166.
}

\lref\BertoldiSF{
  G.~Bertoldi, F.~Bigazzi, A.~L.~Cotrone and J.~D.~Edelstein,
  ``Holography and Unquenched Quark-Gluon Plasmas,''
  arXiv:hep-th/0702225.
}

\lref\SrinivasanTY{
  K.~Srinivasan and T.~Padmanabhan,
  ``Particle production and complex path analysis,''
  Phys.\ Rev.\  D {\bf 60}, 024007 (1999)
  [arXiv:gr-qc/9812028].
}

\lref\CotroneQA{
  A.~L.~Cotrone, J.~M.~Pons and P.~Talavera,
  ``Notes on a SQCD-like plasma dual and holographic renormalization,''
  arXiv:0706.2766 [hep-th].
}

\lref\PreskillTC{
  J.~Preskill,
  ``Do black holes destroy information?,''
  arXiv:hep-th/9209058.
}

\lref\SeibergZK{
  N.~Seiberg,
  ``New theories in six dimensions and matrix description of M-theory on  T**5
  and T**5/Z(2),''
  Phys.\ Lett.\  B {\bf 408}, 98 (1997)
  [arXiv:hep-th/9705221].
}

\lref\EllisJZ{
  J.~R.~Ellis, J.~S.~Hagelin, D.~V.~Nanopoulos and M.~Srednicki,
  ``Search For Violations Of Quantum Mechanics,''
  Nucl.\ Phys.\  B {\bf 241}, 381 (1984).
}

\lref\tHooftRE{
  G.~'t Hooft,
  ``On The Quantum Structure Of A Black Hole,''
  Nucl.\ Phys.\  B {\bf 256}, 727 (1985).
}

\lref\KrausFH{
  P.~Kraus and F.~Wilczek,
  ``A Simple stationary line element for the Schwarzschild Geometry, and some
  applications,''
  arXiv:gr-qc/9406042.
}

\lref\KrausBY{
  P.~Kraus and F.~Wilczek,
  ``Selfinteraction correction to black hole radiance,''
  Nucl.\ Phys.\  B {\bf 433}, 403 (1995)
  [arXiv:gr-qc/9408003].
}

\lref\ParikhQH{
  M.~K.~Parikh,
  ``New coordinates for de Sitter space and de Sitter radiation,''
  Phys.\ Lett.\  B {\bf 546}, 189 (2002)
  [arXiv:hep-th/0204107].
}

\lref\WaldKC{
  R.~M.~Wald,
  ``On Particle Creation By Black Holes,''
  Commun.\ Math.\ Phys.\  {\bf 45}, 9 (1975).
}

\lref\Zurek{
  W.~H.~Zurek,
  ``Entropy evaporated by a black hole,''
  Phys.\ Rev.\ Lett.\  {\bf 49}, 1683 (1982).
}

\lref\BraunsteinSJ{
  S.~L.~Braunstein and A.~K.~Pati,
  ``Quantum information cannot be completely hidden in correlations:
  Implications for the black-hole information paradox,''
  Phys.\ Rev.\ Lett.\  {\bf 98}, 080502 (2007)
  [arXiv:gr-qc/0603046].
}

\lref\BarbonZA{
  J.~L.~F.~Barbon, C.~A.~Fuertes and E.~Rabinovici,
  ``Deconstructing the Little Hagedorn Holography,''
  arXiv:0707.1158 [hep-th].
}

\lref\HemmingAS{
  S.~Hemming and E.~Keski-Vakkuri,
  ``Hawking radiation from AdS black holes,''
  Phys.\ Rev.\  D {\bf 64}, 044006 (2001)
  [arXiv:gr-qc/0005115].
}

\lref\WittenZW{
  E.~Witten,
  ``Anti-de Sitter space, thermal phase transition, and confinement in  gauge
  theories,''
  Adv.\ Theor.\ Math.\ Phys.\  {\bf 2}, 505 (1998)
  [arXiv:hep-th/9803131].
}

\lref\DineKT{
  M.~Dine, O.~Lechtenfeld, B.~Sakita, W.~Fischler and J.~Polchinski,
  ``Baryon number violation at high temperature in the standard model,''
  Nucl.\ Phys.\  B {\bf 342}, 381 (1990).
}

\lref\BekensteinHC{
  J.~D.~Bekenstein,
  ``Nonexistence of baryon number for static black holes,''
  Phys.\ Rev.\  D {\bf 5}, 1239 (1972).
}

\lref\GiveonDK{
  A.~Giveon and D.~Kutasov,
  ``The charged black hole / string transition,''
  JHEP {\bf 0601}, 120 (2006)
  [arXiv:hep-th/0510211].
}

\lref\AtickSI{
  J.~J.~Atick and E.~Witten,
  ``The Hagedorn Transition and the Number of Degrees of Freedom of String
  Theory,''
  Nucl.\ Phys.\  B {\bf 310}, 291 (1988).
}

\lref\SusskindWS{
  L.~Susskind,
  ``Some speculations about black hole entropy in string theory,''
  arXiv:hep-th/9309145.
}

\lref\HottaYJ{
  K.~Hotta,
  ``The information loss problem of black hole and the first order phase
  transition in string theory,''
  Prog.\ Theor.\ Phys.\  {\bf 99}, 427 (1998)
  [arXiv:hep-th/9705100].
}

\lref\PonsKY{
  J.~M.~Pons and P.~Talavera,
  ``Truncations driven by constraints: Consistency and conditions for  correct
  upliftings,''
  Nucl.\ Phys.\  B {\bf 703}, 537 (2004)
  [arXiv:hep-th/0401162].
}

\lref\SusskindMU{
  L.~Susskind and L.~Thorlacius,
  ``Gedanken experiments involving black holes,''
  Phys.\ Rev.\  D {\bf 49}, 966 (1994)
  [arXiv:hep-th/9308100].
}

\lref\thorne{
K.~S.~Thorne, R.~H.~Price and D.~A.~MacDonald,
``Black Holes: The membrane paradigm,'' Yale University Press,
New Haven, CT, (1986).
}

\lref\SusskindIF{
  L.~Susskind, L.~Thorlacius and J.~Uglum,
  ``The Stretched Horizon And Black Hole Complementarity,''
  Phys.\ Rev.\  D {\bf 48}, 3743 (1993)
  [arXiv:hep-th/9306069].
}

\lref\DeoBV{
  N.~Deo, S.~Jain and C.~I.~Tan,
  ``String Statistical Mechanics Above Hagedorn Energy Density,''
  Phys.\ Rev.\  D {\bf 40}, 2626 (1989).
}

\lref\HerzogRA{
  C.~P.~Herzog,
  ``A holographic prediction of the deconfinement temperature,''
  Phys.\ Rev.\ Lett.\  {\bf 98}, 091601 (2007)
  [arXiv:hep-th/0608151].
}

\newsec{Motivations}
\seclab\intro

A central theme in the black hole information puzzle is the problem of low-energy scattering for ordinary quanta by an extremal black hole with a subsequent absorption and Hawking reemission. From 
a semi-classical point of view the final radiation turns to be that of an exact black body \refs{\HawkingRA,\WaldKC}.
It has been argued, but not demostrated, that departures
from thermal emission could explain black hole evaporation 
without lost of information and hence reconcile quantum mechanics 
with general relativity.
In most of the approaches in the literature 
the role of the black hole is
similar to that of a soliton in field theory, being 
gravity treated as a 
non-perturbative field to be added to the game once the
spectrum and quantization rules to the particle-like objects
have been put down by quantum mechanic rules. Although this view
must suffice in a semi-classical picture it can be inappropriate when one 
probes Planck scales.

One successful approach that overcomes partially this problem,
incorporates the self-gravitation interaction in
the radiation process \KrausBY.
The underlying idea in this model is extremely simple: the full hole-particle system is reduced to an effective one-dimensional system and for that purpose all the degrees of freedom are truncated to 2d. 
In particular the model for emission/absorption is still
only suitable for regions of low-curvature and exclusively tackles the s-wave part of the short-wavelenghts radiation. This fact allows to employ the WKB approximation that makes any calculation almost straightforward.
{\sl All} the studies pursued within the mentioned approach
reveal so far that Hawking radiation is not purely
thermal. These results, although encoraging to explain the Hawking 
effect, are distressing and is not clear the ultimate reason
that allows to {\sl all} the holes to have a non-thermal emission independently of their nature. 
Our aim is to present, some features of the semi-classical geometry and Hawking radiation
in a family of black holes with strict thermal emission even if back-reaction effects are taken into account. 

We shall begin by outlining the most salient features of a simpler related model, Little String Theory (LST), that is at the main core of the study. Many of the points that will arise here are implicit or explicit given in other works.
Next we present the emission probability via tunneling in this model, 
explaining some details of the formalism. As a next step we elucidate a plausible ``dynamics''  that rides a NS5 setup towards its Hagedorn temperature and study the spectrum of the emission. As we shall see as temperature is increased in this process the spectrum, initially non-thermal, goes to a thermal one.

To stress that the thermal emission is not something 
peculiar of this metric space, but most probably a feature
of a full
family of spaces \GubserEG, we also worked out  a model which ultraviolet completion reduces to the previous one. In that sense one does not expect to obtain the very similar
result as before for the decay width, because the emission/absorption process is 
produced near the horizon and must be insensible to the behaviour of the radial asymptotic in the metric. 
As we shall see this does not turn to be the case. 

To conclude we add a few remarks on the information lost and higher order corrections near the Planck scale. 

\newsec{Little String Theory, thermodynamics overview}
\seclab\lst

The model we
study is constructed by considering $N$ coincident NS5-branes in type II string 
theory in the limit of a vanishing asymptotic value for the string coupling $g_s\to 0$ and a  
fixed string mass $m_s$.  Under these constraints the theory becomes free in the bulk but strongly interacting on the brane,
while modes interacting between the bulk and the brane are decoupled. This setup defines a non-gravitational six-dimensional field theory \SeibergZK. 
In that precise limit the theory reduces to LST, or more precisely to $(2,0)$~ LST for type IIA NS5-branes and 
$(1,1)$~ LST for type IIB NS5-branes \AharonyUB. 

We shall consider the non-extremal case, from where we shall deduce the thermodynamic properties of the black hole. 
Even if the Hawking's area theorem applies in Einstein frame, where the weak energy condition is satisfied \HawkingTU, we have cross-checked that all our claims concerning the semi-classical emission can also be obtained from the String frame where from simplicity we stick henceforth.
The classical throat geometry corresponding to $N$ coincident non-extremal NS5 branes is 
 described by \CallanAT,
\eqn\lstgeo{
ds^2 = - F(r)
dx_1^2+ \sum_{j=2}^6 dx_j^2+{N\over m_s^2 r^2}  \left( {dr^2\over F(r) }
+ r^2 d\Omega_3^2\right)\,,\quad  F(r) = 1-{r_0^2\over r^2} \,,}
where the dilaton field is given by
$e^{2 \phi}= {N\over m_s^2 r^2}\,.$
The boundary of the near horizon geometry is $R^5\times S^1\times S^3$ and only reduces to $R^5$ after
Kaluza-Klein reduction on the $S^1$ and $S^3$ spheres. 

The extremal configuration is obtained by identifying $F(r)\to 1$ in \lstgeo. This
represents a five-brane which world-volumen can be identified with ${\cal R}^6$.
In addition to the previous
fields one finds a NS-NS $H_{(3)}$ form along the $S^3$, $H_{(3)}= 2 N_c \epsilon_3$. 
The geometry \lstgeo~is regular as
long as $r_0\ne 0$. When
$r$ approaches $r_0$, appears a semi-infinite ``throat'' parametrized by the $(x_1, r)$ 
coordinates.
The dilaton field grows linearly in this region, pointing out that gravity becomes strongly coupled far down the throat.
As we shall see in this geometry
there are null infinities, since light rays can travel forever down this throat. 

{}From the point of view of the black hole thermodynamics,
the thermal states are constructed by periodically identifying the imaginary time coordinate $x_1$ with
a period 
\eqn\temp{
 \beta_0= {2 \pi \sqrt{N}\over m_s}\,.
}
Notice that this value is independent of the black hole radius, that is {\sl fixed}
even if many particles impinge on the black hole. Furthermore, eq. \temp\ gives
the onset for the characteristic time scale, the so-called Hawking time $\tau_{\rm H}$, in which the black hole is formed 
$\tau_{\rm H} = \sqrt{N}/\left(2 m_s\right)$.
These thermodynamic states will be in thermal equilibrium in the static coordinate system \lstgeo\  with a locally measured temperature
\eqn\localtemp{
T_{\rm loc}(r) =  {1\over \beta\sqrt{F(r)}} \,.
}
The local temperature is blue-shifted by the gravitational potential and increases as $(r-r_0)^{-1/2}$
for $r \to r_0$. An asymptotic observer will identify its observed temperature with that in \temp, $T_{\rm loc}(r\to\infty) = \beta_0^{-1}$. Thus although
the black hole has a natural, fixed, temperature associated with it, in this case the locally measured temperature decreases, up to $\beta_0^{-1}$, the further one is from the black hole.
\medskip
It has been argued, \WittenZW, that the energy, entropy and temperature of a CFT at high temperatures
can be identify with the mass, entropy and Hawking temperature of the dual black hole and in the sequel we shall make use of these relations.
The Euclidean action for a LST black hole solution gives a vanishing contribution to the Helmholtz free energy $\log Z = - {\cal I}=0$, with $Z$ been the string partition function. In that precise case the entropy and energy density are directly proportional to each other, 
\eqn\enten{
s= \beta_0 e= {\pi^2\over 2} \sqrt{N} r_0^2\,, 
}
and the Bekenstein-Hawking entropy-area relation is fulfilled.
This behaviour suggest
that at leading order the Hagedorn density of states at very high energy grows as 
$\rho(E) \sim exp\left(\beta_0 E\right)$  \AharonyTT.
As a consequence the energy of strings near the Hagedorn temperature is dominated by the oscillating mode energy, i.e. the mass energy of a single string.

\newsec{Hawking emission via tunneling}
\seclab\em

We give in this section a somewhat detailed derivation for the obtention of the Hawking radiation. 
In the subsequent sections, sec. 4 and 7, we shall take the same approach but omitting details and commenting directly on the results.

 \bigskip
 
Following \ParikhMF~we consider the emission of a S-wave massless scalar particle
in the radial direction of \lstgeo. This will allow to use Birkhoff's theorem and decouple gravity from matter.
In order to find the Hawking emission we bring the length element \lstgeo~ to a smooth
form near the horizon using a Painlev\'e-like
transformation $x_1\to \hat{x}_1+f(r)$, which is nothing more than the proper time along the radial geodesic worldline \KrausFH.
This form will be more suitable to study across-horizon physics, for instance the tunneling of
massless shells.
In doing so, we consider a transformation with the property
that at a constant time slice matches the geometry of LST space 
without a black hole immersion
\eqn\nobh{ 
ds^2 =  \sum_{j=2}^6 dx_j^2+N \left( {dr^2\over r^2 }
+d\Omega_3^2\right)\,.}
This is acomplished by choosing 
\eqn\derf{
f(r) = -\sqrt{N}\, {\rm arctanh}\left({r\over r_0}\right)\,,
}
which allows to rewrite \lstgeo~ as
\eqn\lstp{
ds^2 = - F(r)
d\hat{x}_1^2+ \sum_{j=2}^6 dx_j^2-2 \sqrt{N}{r_0 \over r^2} dr
 d\hat{x}_1 + 
{N\over r^2} \left( dr^2
+r^2 d\Omega_3^2\right)\,.}
The function \derf\ is time independent and as a consequence
\lstp\ remains stationary as was already the case for \lstgeo.

To describe the black hole emission we rely on the notion of  
virtual pair creation just around the  
horizon \HawkingSW. Loosely speaking, if the pair is created inside the horizon
the positive energy particle tunnels out
while the antiparticle is absorbed by the black hole which horizon recesses.
Alternatively the pair can be created just outside the horizon, 
in that case is the
antiparticle which tunnels throught the horizon, shrinking once more the size of the
black hole while the particle escapes. In any of the cases the quantum state of the outside particle is not a pure state, and it is possible to compute the entanglement entropy between the particles that fall into the hole with those that escape to infinity.  

This intuitive picture contains some drawbacks, the main one being the lack of understanding on the origin of the source for the potential barrier to tunnel across.  The approach devised in \refs{\ParikhMF,\ParikhIH}~overcome this by noticing that when a virtual pair of particles is created is the self-gravitating field of the emitted particle the source for the potential barrier to tunnel across the horizon. 
In addition one has to take into account the energy conservation in the process: the ADM mass
remains fixed while the black hole mass decreases when the quanta is emitted.
This backreaction deforms the initial metric and is implemented in \lstgeo\ by shifting the black hole mass appearing in the wrapping factors, $M\sim r_0^2$.
To be concrete, once the shell is emitted
the correct wrap factor would be proportional to $M-\omega$, with $\omega$~been the energy released
in the emission. This would correspond to a new, lower value for the radius $r_1$.

{}For an observer located at the 
radial 
infinity of \lstgeo, an object approaching $r_0$ is infinitely blueshifted. This allows
to apply a semi-classical treatment to the particle emission problem and with an extend 
to use the classical
action, in the smooth coordinates \lstp, to describe the wave function
$\Psi(r)\sim e^{i S_{\rm class}}\,.$ 
Keeping this in mind we evaluate the rate emission for massless particles in the sequel.

The metric \lstgeo~ is stationary and the lagrangean density derived from it
fulfills the simple 
relation
${\cal H} = -2 {\cal L}$~with the hamiltonian density. For a dynamics considering 
only the radial coordinate the expression 
${\cal L} = - {\dot r} p_r$~ holds 
and the classical action
reads as
\eqn\ac{
S= \int_{r_{\rm in}}^{r_{\rm out}} p_r dr =
\int_{r_{\rm in}}^{r_{\rm out}}
\int_{M}^{M-\omega} {d H \over \dot{r}} dr =
- \int_0^\omega d\omega\int_{r_{in}}^{r_{out}}{dr\over {\dot r}}\,,
} 
being $\omega$ the maximum energy released in the shell. To obtain \ac\
we have applied Hamilton's equation, defined $\dot{r} :={dr/d\hat{x}_1}$ and pulls out
factors that do not contribute to the imaginary part of the action. 
Inherently the expression \ac\ is obtained in the semi-classical regimen, i.e. 
the emitted shell must be a {\sl probe}, $\omega\ll {\rm M}$.
This also is justified because for large holes masses, much larger than Planck mass, the only relevant field configurations taken into account by the WKB approximation are short wavelength solutions in a relative low curvature region. This in addition
overcomes the ill-defined extremal limit  \PreskillTB.

{}For the geometry \lstp\ 
the radial light-like geodesic are orthogonal  to the surfaces of constant time 
on which $r$ measures the radial proper distance and is given by
\eqn\geo{
\dot{r} = {1\over \sqrt{N}} ( r \pm r_0)\,,
}
where the plus (minus) sign corresponds to the geodesics rays going towards
(away from) the observer.
Its general solution is $ r= r_0 \left(e^{\hat{x}_1/\sqrt{N}}
\pm 1 \right)\,.$ As mentioned in 
sec. \lst  any radial light-like emission reach
future null infinity at $\hat{x}_1\to\infty$. While a light-like emission leaving the observer at $\hat{x}_1=0$~reach the horizon at $\hat{x}_1=\ln 2/\sqrt{N}\,,$ thus eventually as one increases the number of NS5-branes the traveling time gets reduced. 

Using the Feynman prescription
$+ i \epsilon$ 
to displace the pole, the imaginary part of \ac\ reads
$
{\rm Im S} =  \pi \sqrt{N} \omega\,.
$
One does not fail to notice that: {\sl i)}  this result is independent of the black hole radius
and {\sl ii)} 
that no infinities arise in this calculation, so is mathematically well defined 
without any need for regularization. 
The previous relation, together with \temp, leads to the rate emission
\eqn\emi{
\Gamma  \sim \vert \Psi(r) \vert^2  \sim e^{-\beta_0 \omega}\,.
}
The exponent contains the difference between the actions of the higher and lower black hole mass evaluated at
the same and unique temperature for the system. The emission \emi\ follows a black body distribution
and hence the LST black hole radiation is pure {\sl thermal}. 

The consequences of \emi\ are: {\sl i)} first of all that all the corresponding states in the dual CFT must be a priori equally weighted. {\sl ii)} Secondly, one can convince oneself that cluster decomposition applies  and as a result
the quantum state of Hawking radiation does not depend on the initial state of the collapsing body.
In addition this fact implies that the
probability of emission of a shell of energy $\omega_1+\omega_2$ is equal to the probability of emitting independently two shells with the same total amount of energy.

As the radiation comes always as a pure state, the Hilbert space can be factorized into two disjoint parts, ${\cal H} = {\cal H}_{\rm in} \oplus{\cal H}_{\rm out}$, which correspond to states located at the inner and outer sides of the event horizon respectively. It will follow from the 
superposition principle that the state inside the horizon must 
be a unique state carrying no information at all.
Summing up, this can be expressed in a somewhat muted fashion as: the black hole at the hagedorn temperature does not interact with its environment and hence
we can represent a state of the entire space as $\vert \psi(t) \rangle = \vert \psi_{\rm in}(t) \rangle\otimes \vert \psi_{\rm out}(t) \rangle$.  

Momentally we made a digression of our main stream and comment on the validity of the truncation of \lstgeo\ to 2-dimensions. The interesting points concern: {\sl i)} the fate of dimensional and field content reduction on the $S^3$ modes is consistent \PonsKY. {\sl ii)} Furthermore, both the $R^5$ and $S^3$ wrap factors are independent of the $(\hat{x}_1,r)$ coordinates. As a consequence the equation of motions of these modes can be taken static and $r$ independent, i.e. the emission in the $\hat{x}_1-r$ plane does not alter the dynamics in the transverse coordinates to it.

\newsec{Locking information at the Hagedorn temperature}

That the result \emi\ must be the correct behaviour for the LST system 
is intuitively clear in the semi-classical approach
from the very beginning because in this type of holes the
temperature is not related with its mass. It is precisely
this fact which encodes the ultimate reason for the non-thermal behaviour in the model of \ParikhMF. 
To make this point more clear if instead of using the field content of LST we retain the full asymptotic, 
ten-dimensional CHS background \CallanHS
\eqn\chs{
ds^2 = - F(r)
dx_1^2+ \sum_{j=2}^6 dx_j^2+A(r)  \left( {dr^2\over F(r) }
+ r^2 d\Omega_3^2\right)\,,\,\, A(r)=\kappa+{ N\over m_s^2 r^2}\,, 
}
and dilaton
$e^{2 \phi}= \kappa+{ N\over m_s^2 r^2}\,, (\kappa\equiv 1)\,,$ one sees that the temperature depends on the black hole mass  \BarbonZA. In this case the Hawking temperature can be determined by the surface gravity method at the event horizon and is given by
\eqn\bechs{
\beta_{\rm CHS} = \beta_0 \sqrt{1+\kappa r_0^2/ N}\,,
}
notice that it provides an infra-red cutoff for the radial coordinate.
We have used $\kappa$ as an eventual continuos variable
that parameterizes the geometry \chs.  By no means, one should not understand that all the intermediate values correspond to supergravity solutions. Its utility is twofold, first the near horizon limit
is recovered setting $\kappa =0$. And second it will also control the temperature; for instance
$\kappa \to 0$ increases the temperature to the Hagedorn one. 
The basic tenant is that \bechs\ relates the temperature with the size of the hole, thus
as the hole emits, not only the radius shrinks but also the temperature increases. This fact relates the emission with the thermodynamic properties of the hole and contrary to the previous situation we expect that the radiation
provides information on the black hole state.

As previously the geometry at the horizon can be brought to a smooth form with a Painlev\'e-like change of coordinates 
\eqn\derfchs{x_1\to
\hat{x}_1-r \sqrt{ A(r) -\kappa F(r) }\, {\rm arctanh}\left( {r\over r_0} \sqrt{ 1 -\kappa {F(r) \over A(r)} }\right)
+ r_0 \sqrt{A(r)} \log\left[ 2 r \left(\sqrt{\kappa}+\sqrt{A(r)}\right)\right].
}
After using \derfchs\  the metric field \chs\ is reduced to
\eqn\chspain{
ds^2 = - F(r)
d\hat{x}_1^2+ \sum_{j=2}^6 dx_j^2-2 \sqrt{A(r)}\, {r_0\over r} d\hat{x}_1 dr +A(r)  \left( dr^2\
+ r^2 d\Omega_3^2\right)\,.
}
A calculation similar to \ac\ leads to the probability for a CHS black hole of mass $M$ to emit a shell of energy $\omega$ 
\eqn\emichs{
\Gamma\sim \exp\left(-2 \pi  \sqrt{N + M \kappa}\,\,\omega  + {\kappa\, \omega^2\over 4 \sqrt{N+M \kappa}}+\ldots \right)\,,
}
where the ellipsis stand for terms proportional to higher powers of $\kappa$.  Now for $\kappa\to 1$ \emichs\
is clearly non-thermal while for $\kappa\to 0$ we recover once more the thermal emission \emi.
In view of this fact it seems wholly tenable that as the temperature is increased, $\beta_{\rm CHS} \to \beta_0$, the system evolves from been non-thermal to be thermal, and as a consequence an asymptotic observer
could conjecture that the black hole internal degrees of freedom are reduced during the evaporation process
and eventually one remains with a single state. 
The very same conclusions can be traced back from a stringy point of view if one consider the strings as the fundamental degrees of freedom of the black hole. In a flimsy language: as one approaches the Hagedorn temperature strings condense leaving a residual single one, a unique state that contains no information at all \DeoBV.
 To substantiate this point we have computed, in the spirit of \SusskindIF, some properties of
a classical string located at the stretched horizon, i.e. a time-like curve slightly outside the global event horizon,  that is of relevance in describing the evaporation process.
We expect that for sufficiently large black hole masses both the proper distance between the stretched  and  the event horizon, $\sim \int_{\rm e.h.}^{\rm s.h.} dr \sqrt{g_{rr}}$, together with 
the local Unruh temperature,  \localtemp, are ballpark of the Planck order (up to a numerical factor of order $1$). This imposes that the stretched horizon must be almost coincident with the event horizon, $r_p \approx r_0 + \delta$ for some positive and {\sl infinitesimal } constant $\delta$.
Using \localtemp\ at the Planck radius and the Planck temperature behaviour, $T_p\sim G^{-1/2}$, we obtain
\eqn\del{
\delta \approx { G \sqrt{G M}\over   \beta_0^{2}+4 G M \kappa}\,,
}
where we have momentally reinstated the Newton constant $G$ in the proper space-time dimension.
{}For the CHS model $\delta \sim \sqrt{ G/ M}$, thus for large black hole masses one can consider that the stretched horizon is almost on top of the event horizon. As we increase the temperature the distance $\delta$ also increases up to reaching $\delta \sim G \sqrt{G M}/\beta_0^2$ at the Hagedorn temperature. At this point the stretched horizon is displaced towards the distant observer and swallows up all of space, provided we ensure the validity of the supergravity approximation
\eqn\sugval{
M\sim r_0^2 \gg N \gg 1\,. 
}
In the CHS model all the thermodynamic quantities on the stretched horizon can be identify as those of the event horizon, with additional subleading terms suppressed by the hole mass. This is in contrast with the outcome at the Hagedorn temperature where subleading contributions are not longer suppressed.

Let us continue examining the classical behaviour of the stretched horizon and visualize
the ``number of states''. For that purpose we calculate,
in the 2-dimensional flat Minkowsky space, the mass of a ring shaped string located between the boundary at the Planck temperature, $T_p$, and the event horizon. It reads
\eqn\stmass{
m=\int_{\sqrt{G M}}^{ \sqrt{G M} + \delta} 2 \pi r \rho_p\, dr \approx \left\{
\matrix{
{1\over G M}\,, &{\rm if}\,\, \kappa=1;  \cr
& \cr
{ M \over \beta_0^2}+ {\cal O}\left({G M \over \beta_0^4}\right) \,, &{\rm if}\,\, \kappa=0 \cr}  \right. 
}
where we have used the behaviour $\rho_p\sim G^{-2}$. 
Notice that \stmass\ matches the speculations below \emichs: for the background \chs\ the string mass can be considered residual and in accordance the black hole mass remains to be almost $\sim G M$.
Furthermore, the whole mass is localized inside the event horizon.
As we increase the temperature the mass of the string forming a ring of radius $r_p$ is of the order of the black hole mass and hence there must be only a residual mass in the interior of the event horizon. With the expectation of a small distortion w.r.t. the flat Minkowsky space the approach
 of \stmass\ is fully justified in this latter case.
One can regard this phenomenon as a progressive melting of the strings as they encounter Hagedorn temperature conditions \AtickSI. The energy of the strings states is so large when the Hagedorn temperature is approached, that strings on the horizon will tend to join forming a single one \SusskindWS. 
Thus the system evolves to a single state and consequently the entropy is reduced.
This picture matches the view where black hole states at the Hagedorn temperature are in one to one correspondence with single string states. 

\newsec{Validity of the Semi-classical approach}
\seclab\valsc

The previous analyses are based on a semi-classical approach, and even if in top of them one can implement some extra quantum corrections, the approach is not free of assumptions and possible criticisms. 
{}For instance an observable effect of string theory is the very last steps in the black hole evaporation.
In the usual picture the final evaporation process takes place at planckian temperatures and thus the
last radiated particles would carry energy of order of the Planck scale. One wonders if at this energies
the approach of sec. \em\ is still reliable. If it does, energy conservation imposes a constrain in the minimum size of the remnant, because the energy of the emitted particles can not exceed the remainder mass. 

Common lore assigns to the previous optical approximation treatment a validity meanwhile the 
wavelength of the bulk probe is much smaller than the local curvature of spacetime
\eqn\mom{
{1\over {\rm momentum \,\,scale}}\ll {\rm local\,\, curvature\,\, length\,\, scale}\,.
}
In terms of local coordinates, the curvature length scale, $\Delta r$, can be written as a function of the scalar curvature as $ \Delta r = 1/ \left( g_{rr} \sqrt{{\cal R}} \right)$.  This function is bounded  from below
with a single minimum located at $r\approx r_0/2$, and then \mom\ leads to $P \gg 2/r_0$. As the black hole emits and shrinks, the momenta of the space-like geodesics probe must increase to fulfill the inequality \mom.  At some point
the mass of the emitted probe would be larger than the remaining mass in the hole and the semi-classical approach will break down.

Considering the behaviour of the radial momenta
$p_r \sim p_0 g_{rr} \dot{r} = \omega {\sqrt{N+\kappa r^2}\over r-r_0}$
as a function of the emitted particle energy, we can see that inequality \mom\ leads to
\eqn\inequality{
\omega \gg {\sqrt{2N(3Nr^2+r_0^2(2N+5\kappa r^2)} \over r(r-r_0)(N+\kappa r^2)}\,.
}
Notice that a particle near the horizon needs a large amount of energy in order to escape up to the boundary.

\newsec{Further thermodynamic relations}

One should keep in mind that any observable quantity is computed at the boundary and receives contributions from both supergravity solutions \lstgeo\ and \chs. Usually in a given thermodynamic regimen one solution dominates over the other and most of the bulk of the physical quantity can be computed by considering only one of them. We shall see in the sequel that this is not the case for these models.

The basic thermodynamic quantity at play is the Helmholtz free energy, that  can be casted in terms of the action via the relation ${\cal F}={\cal I}/\beta$.
The action consists of two terms 
\eqn\ac{
{\cal I}={\cal I}_{\rm grav}+{\cal I}_{\rm surf}\,.
}
The former given by 
\eqn\action{
{\cal I}_{\rm grav}= {1\over  2 \kappa^2_{10} } \int_{\cal M} d^{10}x \sqrt{g} 
\left( R-{1\over 2} \partial_\mu \phi \partial^\mu \phi  -{1\over 12} e^{-\phi}
 H_{(3)}^2\right)\,,
}
being ${\cal M}$\ a ten-volume.
And the latter being the surface contribution
\eqn\sur{
{\cal I}_{\rm surf}= {1\over \kappa_{10}^2} \oint_\Sigma K d\Sigma\,,
}
with $\Sigma$ the boundary that encloses the ten-volume ${\cal M}$ in \action.  $K$ is the extrinsic curvature, $K_{\mu\nu}= n^\sigma \partial_\sigma g_{\mu\nu}$ and $n^\sigma \partial_\sigma$ the outward
directed unit normal vector.

If one calculates directly the action \ac\ for the solution \chs\ the result turns to be divergent. To regularize the solution we use an ultraviolet cuttof  $\Lambda$ that eventually will tend to infinity. {}Furthermore, we perform a fiducial renormalization, subtracting  a reference background. It seems natural to choose the latter as the corresponding extremal solution. The calculation is lengthly but straightforward:
the on-shell Euclidean actions of the extremal and non-extremal solutions are given by
\eqn\e{
{\cal I}_{\rm e}= { {\rm Vol}(R^5)  {\rm Vol}(S^3)\over 2\kappa_{10}^2} 
\int_0^{\beta^\prime} dt \left[
{3\over 2} \Lambda^2 \left({3N+4\Lambda^2 \kappa\over N+\Lambda^2 \kappa}\right)
-\int_0^\Lambda dr   {N^2 r\over (N+r^2 \kappa)^2}  \right]\,,
}
and
\eqn\ne{ 
{\cal I}_{\rm ne}= 
{ {\rm Vol}(R^5)  {\rm Vol}(S^3) \over 2\kappa_{10}^2}\!\!
\int_0^{\beta_{\rm CHS} (\Lambda)}\!\!\!  dt \left[
{ N( 9 r^2  - 5 r_0^2)+ 4 \kappa r^2 ( 3 r^2  - 2 r_0^2) \over 2 (N+r^2 \kappa)}
-\int_{r_0}^\Lambda \!\!\! dr   { N r ( N-\kappa r_0^2) \over (N+r^2 \kappa)^2 }  
 \right]
}
respectively. At the boundary, $\Lambda \to \infty$, the temperature of both solutions must be the same. {}For this purpose the temporal period in the extremal case is adjusted to be 
$\beta^\prime=\beta_{\rm CHS}(\Lambda) \sqrt{F(\Lambda)}\,.$

{}For fixed, but otherwise  arbitrary $N$ and $r_0$,
we find the renormalized action
$$
{\cal I}=\lim_{\Lambda \to \infty} \left[ {\cal I}_{\rm ne}-{\cal I}_{\rm e}\right]
=\lim_{\Lambda \to \infty}  {1\over 4\kappa_{10}^2}    { {\rm Vol}(R^5) (2 \pi)^3\over (N+\kappa \Lambda^2)^{3/2} }
\left( -2 \Lambda (2 N + 3 \kappa \Lambda^2) (N+\kappa r_0^2) \sqrt{\Lambda^2 -r_0^2} +
\right.
$$
\eqn\I{
+
\left.
N^2 (4 \Lambda^2 - 2 r_0^2) +2 \kappa^2 \Lambda^2 (3 \Lambda^2 -2 r_0^2) r_0^2 + 
N \kappa (6 \Lambda^4 + \Lambda^2 r_0^2 - 3 r_0^4){ \over}  \right) \to 0
}
implying that the free energy of the system vanishes. This means that none of the actions dominate over
the other, and to obtain an observable one has to add the contributions of both actions. 

\medskip

It is also instructive to compute in an independent way some of the thermodynamic 
contributions to the Helmholtz free energy, ${\cal F}= E - T S=0$. {}For instance the entropy
goes as 
\eqn\ent{
S= {{\rm Area}\over 4 G_{10}}=  {1\over 2 G_{10}} {\rm Vol}(R^5) \pi^2 r_0^2 \sqrt{N+\kappa r_0^2}=
 {1\over 4 G_{10}} {\rm Vol}(R^5) \pi r_0^2 \beta_{\rm CHS}\,,
}
and turns to be $\kappa$ dependent, but the combination entering in the Helmholtz free energy it is not 
\eqn\entt{
T_{\rm CHS} S =
{1\over 4 G_{10}} {\rm Vol}(R^5) \pi r_0^2= T_{\rm LST} S\,.
}
Notice that \ent\ matches the behaviour described by \stmass:  as $\kappa \to 0$ the black hole dof, strings, joint together up to forming a single state. As a consequence the entropy decreases.

\medskip

We just end this section by noticing that the exponent in
\emi, the entropy radiation, is just the variation of the Bekenstein-Hawking entropy. In this precise case the mass and entropy density are given in \enten\ from where it
follows that $e^{-\beta_0 \omega} = e^{\Delta S_{\rm BH}}$. This matches the statistical picture in which large fluctuations are suppressed and supports the idea that in this background
the Bekenstein-Hawking area-entropy relation, $S_{BH}= A/4$, can be obtained by counting 
the degeneracy states \StromingerSH.

\newsec{Hawking emission via tunneling: Wrapped fivebranes}

The metric \lstgeo\ is the ultraviolet completion of a large family group of regular non-abelian monopole solutions in ${\cal N}=4$ gauged supergravity,
interpreted as 5-branes wrapped on a shrinking $S^2$ \GubserEG. In the following we shall deal with a thermal deformation of one of such metrics dual to ${\cal N}=1$ SQCD
with a superpotential coupled to adjoint matter \CaseroPT. Analyzing the emission problem with the 
method outlined in sec. \em\ leads to the same result obtained in \temp, i.e. a constant outward flux of particles independent of the black hole characteristics. 
The metric field in Einstein frame is given by
$$
ds^2 = e^{\phi_0 \over 2} r \left[ - K(r) dx_1^2 + \sum_{j=2}^4 dx_j^2 +
+N \alpha^\prime \left( {4\over r^2 K(r)} dr^2 + {1\over \xi} d\Omega_2^2
+{1\over 4-\xi} d\tilde{\Omega}_2^2 \right) \right.$$
\eqn\cnp{\left. + {N \alpha^\prime\over 4} \left( d\psi + \cos\theta d\varphi+
\cos\tilde{\theta}d\tilde{\varphi}\right)^2 \right]\,,
\quad K(r) = 1-\left({r_0\over r}\right)^4\,.
}
In addition we have a dilaton field which is linear $\phi = \phi_0+r$~ and a RR 3-form field.

First of all we truncate the theory to two dimensions, the
radial and temporal one. As previously none of the others play any role
in the emission.
To cast \cnp\ in Painlev\'e coordinates we chose the function 
$f(r)$ in \derf\ as
$
f(r) = \sqrt{N} \log K(r)\,.
$
Then the truncated theory equivalent to \cnp\ is rewritten as
\eqn\cpnp{
ds^2 = 
e^{\phi_0 \over 2} r \left( - K(r) dx_1^2 
+4 N \alpha^\prime {dr^2\over r^2 K(r)} 
- 4 \sqrt{N \alpha^\prime} {r_0^2\over r^3} dx_1 dt
\right)\,.
}

To calculate the semi-classical emission one needs 
the radial null geodesics of the back-reacted metric.
Like the mass scales as $M\sim r_0^4$~ the emission
of a shell with energy $\omega$~ translates in a shift
in the radius, $M-\omega\sim r_1^4$. This leads, after the emission, to the geodesic
\eqn\geocnp{
\dot{r} = {1\over 2 \sqrt{N \alpha^\prime} } r
\left({r_1^2\over r^2} \pm 1 \right)\,.
} 
Its solutions are $r^2=r_1^2\left(e^{\pm  x_1/\sqrt{N \alpha^ \prime}} \mp 1\right)$, and one finds for timings the very same pattern as in the LST case.

Inserting the outgoing solution of \geocnp\ in \ac\ one obtains
$
{\rm Im S} =  \pi \sqrt{N} \omega,
$
from where follows once more the behaviour \emi.
Thus, most probably, all metric which asymptotic completion is LST will emit thermically.

\medskip

As in the LST case one can check that using the
mass density $m=r_0^4 e^{2\phi_0} N^{5/2}$ and 
entropy density $s=r_0^4 e^{2\phi_0} N^2$ \CotroneQA\ the
emission entropy in \emi\ turns to be directly related with
Hawking-Bekenstein entropy, $e^{-\beta_0\omega}=
e^{\Delta S_{\rm BH}}\,.$

\newsec{Remarks and Implications}

We have computed the decay rate for the NS5 and Little String Theory black holes. The latter can be interpreted as the thermal limit of the former. The entire process of black hole evaporation, except for the final period when the black hole is of Planckian size, can be summarized according to the following patterns: 
Starting from the NS5 system at a given temperature we checked, in a semi-classical approximation, that
the black hole emission is non-thermal \emichs. The black hole contains many degrees of freedom couple with its environment.
At this point the system is thermodynamically irreversible, and the entropy of the surrounding increases as the black hole emits.
As the emission takes place the black hole temperature increases while, both the mass and
the emission rate, decreases becoming the latter pure thermal at the Hagedorn temperature \emi. The interference term vanishes at this point and the black hole system is thermodynamically reversible and consists of a single state. This single state radiates, while the hole temperature keeps completely independent of its mass. Thus, as the LST black hole evaporates, its energy flux is exactly constant. 

Once this point is reached, one could think that we deal with a stable remnant with zero entropy.
That this is not the case can be inferred from the stringy correction to the entropy as a function of the energy. This gives a thermodynamically unstable system \KutasovJP\ which in turn implies that the probability of emission diverges.
In order of having a gross idea of the latter effect we use the area law relation but incorporating its first {\sl quantum} corrections
\eqn\arec{
S_{\rm c}= {{\rm Area}\over 4} + \alpha \log\left({{\rm Area}\over 4}\right)+ {\gamma\over {\rm Area}}+\ldots\,.
}
Taking into account the relations of the mass and energy densities, the black hole emission \emi\ is replaced at leading order by
\eqn\newemi{
\Gamma \sim \left({{\rm Area}_1\over {\rm Area}_0}\right)^\alpha e^{\Delta S_{{\rm BH}}}=
\left(1-{\omega \over M}\right)^\alpha e^{-\beta_0 \omega}\,.
}
The above expression together with the fact that the value of 
$\alpha$ is negative --the system is unstable--  shows that the trend in \newemi\ is that as the system evolves in time the emission increases,
i.e. without further considerations at play the system would fully evaporate without leaving any relic behind it.
This fact is clearly driven by the sign of $\alpha$, which is negative, and 
makes the distinction with results of \ArzanoRS, where the width decay vanishes.
Obviously, the above picture relies in a truncation of \arec\ and as one approaches Planck scales
one must consider that  
subleading contributions in \arec\ are enhanced and they wash out any solid conclusion.

\medskip

We also have found that
for theories which their ultraviolet completion is LST, the radiation is also that of a blackbody at a fixed temperature \temp.
This thermal effect can be made present in the dual field theory as the violation of the baryon number \BekensteinHC. Even if at high-energy CP symmetry violation is negligible is well known that at very high-temperature is indeed unsuppressed \DineKT\ conforming our findings. 

\medskip

The emission model we have used 
is closely related with the eikonal approximations and we checked that the 
same result can be obtained by using the Hamilton-Jacobi approach  \SrinivasanTY. 

\medskip

As a final remark, we have explicitly checked that the thermal behaviour found in \emi\ is not related 
with the vanishing of the jet-queching parameter in the very same models \BertoldiSF. Even if appealing, 
the idea of non-interaction between the system and its surrounding seems unconnected from the energy lost of a quark pair inside a quark-gluon plasma as can be seen by computing the jet-quenching parameter as a  function of $\kappa$, $\hat{q}(\kappa)=0$. A more plausible reason for this behaviour 
is the absence of a 
Hawking-Page transition in these systems. As we have shown in sec. 6 the system remains always in the confined phase.

\bigskip

{\bf Acknowledgments}

\smallskip

The work of P.T. was supported in part by
 Ministerio de Educaci\' on y Ciencia under contract PR2006-0495
and the European Community's
Human Potential Programme under contract MRTN-CT-2004-005104
`Constituents, fundamental forces and symmetries of the universe'.

\listrefs

\bye